\documentclass{article}
\usepackage[left=1in,right=1in,top=1in,bottom=1in]{geometry}
\usepackage{cite}
\usepackage{enumitem}

\usepackage{amsmath}%
\usepackage{amsfonts}%
\usepackage{amssymb}%
\usepackage{bm}

\newcommand{\name}[1]{{\mathsf{#1}}}

\newcommand{\union}{{\cup}}
\newcommand{\set}[1]{{\left\{#1\right\}}}

\newcommand{\LL}{{\mathcal{L}}}

\newcommand{\sol}[3]{{\name{Sol}_{#1}\left(#2, #3\right)}}
\newcommand{\bound}[2]{{{B_{#1}}\left(#2\right)}}

\newcommand{\comment}[1]{{}}

\title{Optimal non-adaptive solutions for the counterfeit coin problem}
\author{C. Thach Nguyen\footnotemark[1]}
\date{}

\begin{document}
\maketitle
\renewcommand{\thefootnote}{\fnsymbol{footnote}}
\footnotetext[1]{Quora, inc., email: {\tt ncthach@gmail.com}}
\renewcommand{\thefootnote}{\arabic{footnote}}

\begin{abstract}
We give optimal solutions to all versions of the popular counterfeit coin problem obtained by
varying whether
(i)~we know if the counterfeit coin is heavier or lighter than the genuine ones,
(ii)~we know if the counterfeit coin exists,
(iii)~we have access to additional genuine coins,
and (iv)~we need to determine if the counterfeit coin is heavier or lighter than the genuine ones.
Moreover, our solutions are non-adaptive.
\end{abstract}

\section{Introduction}
\label{sec:intro}
The counterfeit coin problem is a well-known mathematical puzzle.
Given a set of $n$ look-alike coins containing a counterfeit one which is a bit heavier than the
others, the goal is to find the counterfeit coin in $k$ weighings (\emph{trials}) with a balance
scale.
Simple arguments show that this is achieveable if and only if $n \leq 3^k$.

More interesting versions are obtained by varying the amount of information available to us, e.g.
whether we have access to some additional coins that are known to be genuine, and the amount of
information we seek, e.g. whether we would like to know if the counterfeit coin is heavier or
lighter than the counterfeit coin.
In this paper, we will consider the versions obtained by changing the answers to the following four
questions, two of which deals with the amount of information we have and the other deals with
the amount of information we seek.
\begin{itemize}[noitemsep]
    \item[$\name{Q_1}$:] (\emph{weight comparison}) Do we know if the counterfeit coin is heavier or lighter than the genuine ones?
    \item[$\name{Q_2}$:] (\emph{existence}) Do we know if there is a counterfeit coin?
    \item[$\name{Q_3}$:] (\emph{extra coin}) Do we have access to an additional coin that is known to be genuine?
    \item[$\name{Q_4}$:] (\emph{weight determination}) Do we want to know if the counterfeit coin is heavier or lighter than the genuine ones?
\end{itemize}
Since the answer to $\name{Q_4}$ is vacuously ``yes'' when the answer to $\name{Q_1}$ is ``yes'',
the combination of different answers to these questions yields the 12 versions in Table~\ref{tab:main}.

\begin{table}
\begin{center}
\begin{tabular}{| c | c | c | c | c | c | c | }
    \hline
    Name & $\name{Q_1}$ & $\name{Q_2}$ & $\name{Q_3}$ & $\name{Q_4}$ & $B_i(k)$ \\%& $\CC_i$ \\
    \hline
    $\name{P_1}$    & Y & Y & Y & Y & $3^k$              \\%&                \\
    $\name{P_2}$    & Y & Y & N & Y & $3^k$              \\%& 1, 2           \\
    $\name{P_3}$    & Y & N & Y & Y & $3^k - 1$          \\%&                \\
    $\name{P_4}$    & Y & N & N & Y & $3^k - 1$          \\%& 1, 2, 4        \\
    $\name{P_5}$    & N & Y & Y & Y & $(3^k - 1) / 2$    \\%& 1, 2, 3, 4     \\
    $\name{P_6}$    & N & Y & Y & N & $(3^k + 1) / 2$    \\%& 1, 2, 3        \\
    $\name{P_7}$    & N & Y & N & Y & $(3^k - 3) / 2$    \\%& 1, 2, 3, 4, 5  \\
    $\name{P_8}$    & N & Y & N & N & $(3^k - 1) / 2$    \\%& 1, 2, 3, 5     \\
    $\name{P_9}$    & N & N & Y & Y & $(3^k - 1) / 2$    \\%&                \\
    $\name{P_{10}}$ & N & N & Y & N & $(3^k - 1) / 2$    \\%&                \\
    $\name{P_{11}}$ & N & N & N & Y & $(3^k - 3) / 2$    \\%&                \\
    $\name{P_{12}}$ & N & N & N & N & $(3^k - 3) / 2$    \\%&                \\
    \hline
\end{tabular}
\caption{Optimal relationship between $n$ and $k$ for which different variants of the counterfeit
coin problem have solutions.}
\label{tab:main}
\end{center}
\end{table}

A solution to one of these problems includes two pieces of information: (i) the trials' descriptions,
i.e. the set of coins put on each side of the scale in each trial, and
(ii) a map from their results to the answer, i.e. the identity of the counterfeit coin
and whether it is lighter or heavier than the genuine ones if such information is desired.
In an \emph{adaptive} solution, the description of a trial can be affected by the results of the
previous ones.
In a \emph{non-adaptive} solution, the trials are indepdendent.

As an example, consider an instance of $\name{P_1}$ with $9$ coins and $2$ trials where the
counterfeit coin is lighter than the genuine ones.
A simple adaptive solution of this problem is as follows.
In the first trial, we weigh coins $\set{1, 2, 3}$ against $\set{4, 5, 6}$.
The coins weighed in the second trial depend on the result of the first trial:
\begin{itemize}[noitemsep]
    \item If the first set is lighter than the second one, weigh coin $1$ against coin $2$
    \item If the first set is heaver than the second one, weigh coin $4$ against coin $5$
    \item If the scale is balanced in the first trial, weigh coin $7$ against coin $8$
\end{itemize}
On the other hand, a non-adaptive solution to this problem always weighs coins $\set{1, 2, 3}$ against
coins $\set{7, 8, 9}$ in the first trial and coins $\set{1, 4, 7}$ against coins $\set{3, 6, 9}$ in the second trial.
The readers are invited to work out the answer in each the of the possible 9 outcomes of the
trials.

Clearly, non-adaptive solutions can be considered as (elegant) adaptive solutions.
However, it is conceivable that for certain $n$ and $k$, adaptive solutions are achievable but
non-adaptive solutions are not.

In this paper, we show that this is not the case for the stated problems by giving
non-adaptive solutions\footnote{
There are two special cases, namely $P_3$ and $P_4$ with $n = 3^k - 2$, where there is provably no
non-adaptive solution. For these cases, we provide adaptive solutions instead.}
for $n$ and $k$ satisfying $n \leq \bound{i}{k}$, where $\bound{i}{k}$ is given in the sixth column
of Table~\ref{tab:main} and show that when $n > \bound{i}{k}$, no (adaptive) solution exists.

\subsection*{Related works}
This problem has attracted much attention in the mathematic community.
In particular, Dyson~\cite{Dyson46} gave an algorithm and a proof of optimality for $\name{P_5}$ and
claimed (without proofs) the optimal bounds for $\name{P_6}, \name{P_7}$ and $\name{P_8}$.
His algorithm is non-adaptive when $n$ is divisible by 3 and adaptive when it is not.
Born, Hurkens and Woeginger~\cite{BornHW03} provided a non-apdaptive algorithm for all
$n$ for $\name{P_5}$.
Halbeisen~\cite{HalbeisenH95} provided an adaptive solution and optimality proof for
$\name{P_9}$.
%\footnote{It is tempted to consider one of these versions as being ``more general'' than
%others, as the title of~\cite{HalbeisenH95} suggests. Indeed, a solution for, for example,
%$\name{P_7}$ is also a solution for $\name{P_8}$. However, as we shall see, we can solve $\name{P_8}$
%for more coins than we can for $\name{P_7}$ given the same number of trials. Thus, with respect to
%finding the opitmal relationship between $n$ and $k$, each of these versions provides a unique,
%albeit related, challenge.}.

Many other works concern with different extensions of the problem by (i) changing the number of
counterfeit coins (see~\cite{LiuN04, LiuZN05, Bosnjak94, Li94, AignerL97, Pyber86, Tosic83}
and references therein), (ii) changing the weighing device (see~\cite{MarcoK13, Bonis98} and
references therein), (iii) imposing restrictions on the weighings (see~\cite{LiuM06, LiuZn05a}
and references therein), and (iv) looking for other concepts of optimality, e.g. optimal average
number of trials over random instances (see~\cite{WanD97, HuCH94, LinialT82, Mead79} and references therein.)

\subsection*{Remarks}

\subsubsection*{On problems where we don't know if there is a counterfeit coin}
Suppose we do not know if the counterfeit coin, if it exists, is lighter or heavier than
the genuine ones, i.e. the answer to $\name{Q_1}$ is ``no''.
We claim that the problems where we do not know if the counterfeit coin exists,
i.e. the answer to $\name{Q_2}$ is ``no'', are subsumed by the problems where
where we do know if the counter feit coin exits and are required to determine if
it is lighter or heavier than the genuine ones, i.e. the answers to both $\name{Q_2}$ and
$\name{Q_4}$ are ``yes'', in the following sense:
let $\name{P}$ be a problem where the answer to $\name{Q_2}$ is ``no''
and $\name{P'}$ be a problem where the answers to both $\name{Q_2}$ and $\name{Q_4}$ are ``yes'',
then any solution to $\name{P}$ is also a solution to $\name{P'}$
and any solution to $\name{P'}$ can be converted to a solution of $\name{P}$ in a way that
preserves non-adaptiveness.

To this end, consider solutions $\name{S}$ and $\name{S'}$ of $\name{P}$ and $\name{P'}$ respectively.
On one hand, whenever $\name{S}$ claims that a certain coin $i$ is counterfeit on an instance of $\name{P}$, it must
have included that coin in at least one trial because otherwise it cannot distinguish between this
instance and the instance where there is no counterfeit coin.
The result of this trial determines whether coin $i$ is lighter or heavier than the geniune coins.
Thus, $\name{S}$ is also a solution of $\name{P'}$.
On the other hand, whenever $\name{S'}$ claims that a certain coin $i$ is counterfeit on an instance of
$\name{P'}$, it must also have included $i$ in at least one trial because otherwise it cannot tell whether
$i$ is lighter or heavier than the genuine ones.
In other words, $\name{S'}$ never encounters the case where the scale is balanced in all of its trials on instance
of $\name{P'}$.
Hence, we can solve $\name{P}$ by applying $\name{S'}$ and outputing ``no counterfeit coin'' whenever
the scale is balanced in all of its trials.

From this claim, $\name{P_9}$ and $\name{P_{10}}$ are subsumed by $\name{P_5}$ and $\name{P_{11}}$
respectively, and $\name{P_{12}}$ are subsumed by $\name{P_7}$.
Thus, $\name{P_4}$ is the only meaningful problem where the existence of the counterfeit coin is
unknown.

\subsubsection*{On the coins' weights}
If the weights of the counterfeit and genuine coins satisfy some special relationships, e.g. the
counterfeit coin is exactly twice as heavy as the genuine ones are, it may be advantageous to perform
trials where the numbers of coins on the two sides of the scale are different.
Such solutions, however, are not general.
In this paper, we require solutions to work even when the weight difference between a counterfeit
and a genuine coins is so small that the result of a trial
with different numbers of coins on the two side of the scale is entirely predictable -- the side
with fewer coins is always lighter.
With this, we can restrict our attention to solutions that always put the same numbers of
coins on the two sides of the scale in each of its trials.
(Note that the number of coins on the scale can still varies between trials.)

\section{Non-adaptive solutions}
\label{sec:algos}
A non-adaptive solution can be represented by an ordered set of vectors $v_i$ of length $k$
such that $v_{ij}$ is $-1$, $1$ or $0$ if
and only if coin $i$ is on the left side, right side or neither in the $j$-th trial respectively.
When we have access to an extra genuine coin\footnote{As we shall see, having access to
more than one extra coins is equivalent to having access to exactly one extra coin.},
a solution set contains $n+1$ vectors and when we do not, it contains $n$ vectors.

In the following sections, we construct solution sets for the problems in
Table~\ref{tab:main} such that $v_i$ determines the outcomes of the trials when coin $i$ is
counterfeit.
In other words, when coin $i$ is counterfeit and lighter (heavier, respectively) than the genuine
ones, the scale is left-titled (right-titled, respectively), right-titled
(left-titled, repectively) or balanced in the $j$-th trial if and only if $v_{ij}$ is
$1$ ($-1$, respectively), $-1$ ($1$, respectively) or $0$ respectively.
We do so by first specifying sufficient conditions\footnote{In many cases, these conditions
are also necessary.} for the solution sets, then constructing vector sets satisfying them.
The proofs that these conditions are sufficient as well as the constructed sets satisfy them
are usually straightforward and omitted for brevity.
Also for the sake of brevity, we will use $\sol{i}{n}{k}$ to denote the constructed solution set
for $\name{P}_i$ with $n$ coins and $k$ trials.

Finally, observe that solutions for $\name{P_2}$ and $\name{P_4}$ also work
for $\name{P_1}$ and $\name{P_3}$ repsectively.
Thus, we only need to limits our attention to
$\name{P_2}, \name{P_4}, \name{P_5}, \name{P_6}, \name{P_7}$ and $\name{P_8}$.

\subsection{{$\name{P_2}$} and {$\name{P_4}$} (known weight comparison)}
For $\name{P_2}$, we are looking for solution sets such that
\begin{itemize}[noitemsep]
    \item \label{c:1} The sum of all vectors in the solution set is $0^k$.
        A set satisfying this condition is said to be \emph{balanced}.
    \item \label{c:2} The solution does not contain identical vectors.
        Formally, there is no $1 \leq i \neq j \leq n$ such that $v_i = v_j$.
\end{itemize}
The first condition is motivated by the remark about the coin's weights in the previous section.
The second condition is required because if $v_i = v_j$, we could not distinguish between the case
where coin $i$ is counterfeit and teh case where coin $j$ is counterfeit.
%It is readily verify that that with such a solution set, coin $i$ is counterfeit
%if and only if for every trial $j$, the scale is left-titled when $v_{ij} = -1$, right-titled
%when $v_{ij} = 1$ and balanced when $v_{ij} = 0$.
%Thus, these conditions are sufficient for $\name{P_2}$.

Since $\set{-1, 0, 1}^k$ consists of $0^k$ and $(3^k - 1) / 2$ pairs of opposite vectors $(v, -v)$,
we can let $\sol{2}{n}{k}$ contains $n/2$ pairs of opposite vectors if $n$ is even and
$(n-1)/2$ pairs of opposite vectors and $0^k$ if $n$ is odd.

For $\name{P_4}$, we are looking to solution sets that satisfy the above two conditions and does
not contain $0^k$.
The reason we need this extra condition is that if $v_i = 0^k$ then we cannot distinguish between
the case where coin $i$ is counterfeit and the case where there is no counterfeit coin.

When $n$ is even, we can set $\sol{4}{n}{k}$ to be identical to $\sol{2}{n}{k}$, which contains
$n/2$ pairs of opposite vectors.
When $n$ is odd and $n \leq 3^k - 4$, $\sol{4}{n}{k}$ can be constructed by first selecting a set
$A$ of 3 vectors satisfying all the conditions, then removing them
from $\set{-1, 0, 1}^k$ and selecting $(n - 3) / 2$ pairs of opposite vectors in the remaining.
One such set $A$ is $A = \set{(-1, 1) \cdot 0^{k-2}, (0, -1) \cdot 0^{k-2}, (1, 0) \cdot 0^{k-2}}$
where $v \cdot w$ is the concatenation of the two vector $v$ and $w$.

We are left with $\name{P_4}$ with $n = 3^k - 2$.
Unfortunately, there is no non-adaptive solution for this case.
To see this, first note that the above condition are not only sufficient but also necessary.
Next, observe that if a set $X$ sastifies them then $\set{-1, 0, 1}^k \backslash X$
is balanced, does not contains identical vectors and contains $0^k$.
However, there are so such set of size $2$.

For the sake of completeness, we give an adaptive solution instead.
In the first trial, we weigh two sets of $3^{k-1}$ coins against each other, leaving $3^{k-1} - 2$ coins out.
If the scale is not balanced, we get an instance of $\name{P_2}$ with $3^{k-1}$ coins and $k-1$
trials, for which we can apply $\sol{2}{3^{k-1}}{k-1}$.
If the scale is balanced, we get an instance of $\name{P_3}$ with $3^{k-1} - 2$ coins (plus an extra, known-to-be-genuine coin) and $k-1$ trials.
If $k = 2$, this instance is trivial; otherwise, we recurse as if it is an instance of $\name{P_4}$.

\subsection{{$\name{P_7}$} and {$\name{P_8}$} (unknown weight comparison, no extra coin)}
Since a solution to $\name{P_7}$, when it exists, is also a solution to $\name{P_8}$, we
only need to construct a solution set for $\name{P_8}$ with $n = (3^k - 1) / 2$ and
solution sets for $\name{P_7}$ with $n < (3^k - 1) / 2$.
Both constructions will be recursive.

\subsubsection*{Recursive construction of {$\sol{8}{(3^k - 1) / 2}{k}$}}
Besides being balanced and not containing identical vectors, a solution set
for $\name{P_8}$ cannot contain any pair of opposite vectors, i.e. there is no
$1 \leq i \neq j \leq n$ such that $v_i = - v_j$.
The reason for this condition is that if such $i$ and $j$ exist, we would not be able to
distinguish between the case where coin $i$ is counterfeit and heavier than the genuine ones
and the case where coin $j$ is counterfeit and lighter than the genuine ones.
Moreover, to aid the construction, we also require that the
solution set does not contain either $1^k$ or $(-1)^k$.

The base cases of our constructions for $k = 2$ and $k = 3$ are given in Appendix~\ref{app}.
Consider any $k \geq 4$ and let $n = (3^k - 1) / 2$ and $n' = (3^{k-2} - 1) / 2$, we
construct $\sol{8}{n}{k}$ from $\sol{8}{n'}{k-2}$ as follows.
First, we concatenate all vectors in $\LL^2$ to the vectors in $\sol{8}{n'}{k-2}$
and note that the resulting set
$\set{v \cdot w : v \in \sol{8}{(3^k - 1)/2}{k}, w \in \LL^2}$
contains $9n'$ vectors and satisfies all the required conditions.
Since $n = 9n' + 4$, we need to add four more vectors to this set.
Because $\sol{8}{n'}{k-2}$ contains neither $1^k$ nor $(-1)^k$, the vectors
$(-1)^k \cdot (-1, 0)$,
$(-1)^k \cdot (1, 1)$,
$1^k \cdot (-1, 0)$,
and $1^k \cdot (0, 0)$
can be added without violating any conditions.

\subsubsection*{Recursive construction of {$\sol{7}{n}{k}$}}
Besides the conditions required for $\name{P_8}$, we also require that $0^k$ is not in the
solution set.
The reason for this condition is that if $v_i = 0^k$, we would not be able to tell if coin $i$ is
heavier or lighter than the genuine ones if it is counterfeit.

The base case of our constructions for $n < 13$ and $k < 4$ are given in Appendix~\ref{app},
except for $n = 11$ and $k = 3$.
Unfortunately, there is no solution set satisfying all required solutions in this case.
Instead, in Appendix~\ref{app}, we give a solution set that contains $(-1)^k$ but satisifies all
other conditions.
We will make sure that we do not use $n = 11$ and $k = 3$ in our recursion.

For any $n \geq 13$ and $k \geq 4$, consider three cases
\begin{enumerate}
    \item $n \leq (3^{k-1} - 3) / 2$.
    In this case, we have a valid solution set for $n$ coins and $k - 1$ trials.
    Simply adding a 0 to each vector in this solution set give us a solution set for
    $n$ coins and $k$ trials.

    \item $(3^{k-1} - 1) / 2 \leq n < (3^k - 3) / 2$.
    In this case, there exists two numbers $h$ and $\ell$ such that $n = 2h + \ell$,
    $4 \leq h \leq (3^{k-1} - 1) / 2$, $4 \leq \ell < (3^{k-1} - 1) / 2$ and
    $h, \ell \neq 11$.
    Then
    \begin{eqnarray*}
        \sol{7}{n}{k} &=& \set{v \cdot (-1) : v \in \sol{7}{h}{k-1}} \\
            & & \union \set{v \cdot (1) : v \in \sol{7}{h}{k-1}} \\
            & & \union \set{v \cdot (0) : v \in \sol{7}{\ell}{k-1}}
    \end{eqnarray*}

    \item $n = (3^k - 3) / 2$.
    Since $\sol{8}{(3^k - 1)/2}{k}$, as constructed in above, contains $0^k$, we can
    set $\sol{7}{n}{k} = \sol{8}{n + 1}{k} \backslash \set{0^k}$.
\end{enumerate}

\subsection{{$\name{P_5}$} and {$\name{P_6}$} (unknown weight comparison, extra coins)}
Besides being balanced and not containing identical vectors, a solution set for $\name{P_6}$
cannot contain any pair of opposite vectors.
Similar to the case of $\name{P_8}$, if there are $i \neq j$ such that $v_i = -v_j$ then we
cannot distinguish between the case where $i$ is counterfeit and lighter than the genuine ones
and the case where $j$ is counterfeith and heavier than the genuine ones.

For $\name{P_5}$, we also require that $0^k$ is not in the solution set.
Otherwise, if $v_i = 0^k$, we would not be able to determine if coin $i$ is lighter or heavier
than the genuine coins if it is counterfeit.

The solution set for $\name{P_5}$ and $\name{P_6}$ satisfying the above conditions are constructed
as follows.
\begin{itemize}
    \item For $n = (3^k + 1) / 2$, $\sol{6}{n}{k} = \sol{8}{n-1}{k} \union \set{v_n = (-1)^k, v_{n+1} = 1^k}$.
    \item For $n = (3^k - 1) / 2$, $\sol{6}{n}{k} = \sol{8}{n}{k} \union \set{v_{n+1} = 0^k}$.
    Moreover, $\sol{5}{n}{k} = \sol{6}{n + 1}{k} \backslash \set{0^k}$
    since $\sol{6}{n + 1}{k}$, as constructed above, contains $v_{i} = 0^k$ for some $i \leq n$.
    \item For $n < (3^k - 1) / 2$, $\sol{6}{n}{k} = \sol{5}{n}{k} = \sol{7}{n}{k} \union \set{v_{n+1} = 0^k}$
\end{itemize}

\section{Optimality proofs}
We show that fore every $i$, the problem $\name{P}_i$ has no (adaptive) solution if $n >
B_i(n)$.
To this end, note that the trial outcomes can be represented by vectors in
$\set{l, b, r}^k$ where the symbols $l$, $r$ and $b$ denote the left-titled, right-titled and
balanced weighing outcomes respectively.
A solution maps (a subset of) these vectors to sets of configurations where
each configuration consists of (i) the identity of the conterfeit coin and (ii) whether it is heavier
or lighter than the genuine ones.
When the answer to $\name{Q_2}$ is ``yes'', the identity of the counterfeit coin is in the set
$\set{1, 2, \ldots, n}$; when it is ``no'', this identity could also take the value $0$, which means there
is no counterfeit coin.
It is easy to see that for problems where the answer to $\name{Q_4}$ is ``yes'', each outcome vector
is mapped to at most one configuration and for the others, each outcome vector is mapped to at most two
configurations with the same identity of the cointerfeit coin.
Moreover, each configuration must appear in at least one subset in the image of the map.

With this observation, we can establish the upper-bounds $\name{P_1}$, $\name{P_3}$ and $\name{P_5}$
using simple information theoretic arguments.
In addition, since $\name{P_2}$ and $\name{P_4}$ are ``easier'' than $\name{P_1}$ and
$\name{P_3}$ respectively,
the upper-bounds for these problems are also readily established.
Thus, we are left with proving the upper-bounds for $\name{P_6}$, $\name{P_7}$ and $\name{P_8}$.

\subsection{$\name{P_6}$ (extra coins, no weight determination)}
\label{sec:bound_p6}
Consider an outcome vector $c \neq b^k$, then there is some index $i$ such that
$c_i \neq b$.
This means the counterfeit coin exists and is among the one weighed at the $i$-th trial; thus the
outcome of this trial determines whether it is lighter or heavier than the genuine ones.
Therefore, $c$ can be mapped to at most one configuration.
This means the only outcome vector that can be mapped to more than one configurations is $b^k$.

Since there are $2n$ configurations and $3^k$ outcome vectors among which at most one can be mapped
to two configurations, we have
$2n \leq (3^k - 1) + 2$, i.e. $n \leq (3^k + 1) / 2$.

Finally, we note that the above argument works regardless of the number of extra
known to be genuine coins we have access to.
Thus, the upper-bound holds even when we have access to more than one extra coins.

\subsection{$\name{P_8}$ (no extra coin, no weight determination)}
We only need to show that this problem is unsolvable for $n = (3^k + 1) / 2$.
To this end, assume that in the first trial, we weigh $\ell$ coins against
$\ell$ coins, leaving $n - 2\ell$ coins out.
If the scale is balanced, we are left with an instance of $\name{P_6}$
with $n - 2\ell$ coins.
Thus, in other for this to be solvable, we must have
$n - 2\ell \leq (3^{k-1} + 1) / 2$, i.e. $2\ell \geq 3^{k-1}$.
Since $\ell$ is an integer, this means $2\ell \geq 3^{k-1} + 1$.
However, with such $\ell$, if the scale is not balanced, we are left with
$k-1$ trials and at least $3^{k-1} + 1$ possible identities of the counterfeit coin,
which is unsolvable.

\subsection{$\name{P_7}$ (no extra coin, weight determination)}
The argument is almost identical to the previous section, except
that if the scale is balanced in the first trial, we are reduced to
an instance of $\name{P_5}$ with $n - 2\ell$ rings and $k-1$ trials, thus we would
use the upper-bound of $\name{P_5}$ instead of $\name{P_6}$ to bound $\ell$.

\section{Conclusion}
We considered various popular variants of the counterfeit coin problems and resolved them in the
best possible way by constructing non-adaptive solutions and proving matching upper-bound for each
of them.

As mentioned in Section~\ref{sec:intro}, there are many possible extensions of this problem.
Among them, the case where we have more than one counterfeit coin is particularly interesting because
it breaks the nice structure of the information gained after each trial.
One approach to regain some strcture is to prove {\it a-priori} that a non-adaptive solution exits
whenever a solution exits.
However, this approach does not work even for the simplest instance.
Thus, more investigation is required.

\bibliography{weigh}{}
\bibliographystyle{plain}

\appendix
\section{Base cases}
\label{app}
For brevity, we will replace vectors in $\LL^k$ by strings in $\set{l, n, r}^k$, with $-1$, $0$ and $1$
replaced by $l$, $n$ and $r$ respectively.

\begin{itemize}[noitemsep]
\item $\sol{8}{4}{2} = \set{nl, lr, rn, nn}$
\item $\sol{8}{13}{3} = \{llr$, $lnr$, $lrl$, $lrr$, $nll$, $nln$, $nnl$, $nrl$,
    $rln$, $rnn$, $rnr$, $rrn$, $nnn\}$.
\item $\sol{7}{3}{2} = \{nl, lr, rn\}$
\item $\sol{7}{3}{3} = \{nln, lrn, rnn\}$
\item $\sol{7}{4}{3} = \{lrn$, $nln$, $nrl$, $rlr\}$
\item $\sol{7}{5}{3} = \{lrl$, $lrr$, $nln$, $rln$, $rnn\}$
\item $\sol{7}{6}{3} = \{lrn$, $lrr$, $nln$, $nnl$, $rlr$, $rnl\}$
\item $\sol{7}{7}{3} = \{lln$, $lrl$, $lrr$, $nln$, $rnl$, $rnn$, $rnr\}$
\item $\sol{7}{8}{3} = \{lln$, $lrn$, $lrr$, $nln$, $nrl$, $rlr$, $rnl$, $rnn\}$
\item $\sol{7}{9}{3} = \{lln$, $lrn$, $lrr$, $nln$, $nnl$, $nrl$, $rlr$, $rnl$, $rnr\}$
\item $\sol{7}{10}{3} = \{lln$, $lrl$, $lrn$, $nln$, $nlr$, $nnl$, $nrr$, $rnl$, $rnn$, $rnr\}$
\item $\sol{7}{11}{3} = \{lll$, $llr$, $lrl$, $lrr$, $nln$, $nrl$, $nrr$, $rln$, $rnl$, $rnn$, $rnr\}$
\item $\sol{7}{12}{3} = \{lln$, $llr$, $lrl$, $lrr$, $nln$, $nnl$, $nrl$, $nrr$, $rln$, $rnl$, $rnn$, $rnr\}$
\end{itemize}

\end{document}